# ALMA FACTS. II. Large Scale Variations in the $^{12}$CO($J = 2-1$) to $^{12}$CO($J = 1-0$) Line Ratio in Nearby Galaxies

Shinya Komugi[1,2,3], Tsuyoshi Sawada[4,5], Jin Koda[3], Fumi Egusa[6], Fumiya Maeda[7], Akihiko Hirota[4,5], and Amanda M. Lee[3]

[1] Division of Liberal Arts, Kogakuin University, 2665-1 Nakano-cho, Hachioji, Tokyo 192-0015, Japan; skomugi@cc.kogakuin.ac.jp  
[2] National Astronomical Observatory of Japan, 2-21-1 Osawa, Mitaka, Tokyo 181-8588, Japan  
[3] Department of Physics and Astronomy, Stony Brook University, Stony Brook, NY 11794-3800, USA  
[4] National Astronomical Observatory of Japan, Los Abedules 3085 Office 701, Vitacura, Santiago 763 0414, Chile  
[5] Joint ALMA Observatory, Alonso de Córdova 3107, Vitacura, Santiago 763 0355, Chile  
[6] Institute of Astronomy, Graduate School of Science, The University of Tokyo, 2-21-1 Osawa, Mitaka, Tokyo 181-0015, Japan  
[7] Research Center for Physics and Mathematics, Osaka Electro-Communication University, 18-8 Hatsucho, Neyagawa, Osaka, 572-8530, Japan  
Received 2024 November 10; revised 2024 December 26; accepted 2025 January 5; published 2025 February 7

## Abstract

We present $^{12}$CO($J = 1-0$) mapping observations over $\sim 1/2$ of the optical disk of 12 nearby galaxies from the Fundamental CO 1–0 Transition Survey of nearby galaxies (FACTS), using the ALMA Total Power array. Variations in the $^{12}$CO($J = 2 - 1$)/$^{12}$CO($J = 1 - 0$) line ratio $r_{21}$ are investigated. The luminosity-weighted $r_{21}$ of the 11 sample galaxies ranges from 0.52 to 0.69 with an average of 0.61. We use position–velocity diagrams along the major axis and tilted ring models to separate the normal rotating galactic disk from kinematic outliers that deviate from pure circular rotation. We find that $r_{21}$ is systematically higher in outliers compared to the disk. We compare $r_{21}$ between SA, SAB, and SB galaxies, and find no significant difference in the average $r_{21}$ depending on the presence of galactic bars. We find, however, that the radial gradient in $r_{21}$ is bimodal, where a group containing all SA galaxies prefer constant or very shallow $r_{21}$ gradients out to 40% of the optical radius, while another group containing all SB galaxies have a steep $r_{21}$ gradient, decreasing by $\sim$20% before 40% of the optical radius, which also corresponds to the radius of the stellar bar. After this radius, these galaxies become consistent with a constant or shallow trend in $r_{21}$. The large scale trend in $r_{21}$ can have implications for how we interpret observations made solely in the $^{12}$CO($J = 2 - 1$) line.

*Unified Astronomy Thesaurus concepts:* Galaxy disks (589); Molecular gas (1073); CO line emission (262)

## 1. Introduction

Molecular hydrogen gas is a fundamental constituent of galaxies, driving the formation of stars and ultimately its evolution. Due to its lack of a dipole moment, however, it is difficult to directly measure its quantity except in extreme cases of high density and/or temperatures. Carbon Monoxide (CO) is routinely used as a tracer of molecular gas, especially the $^{12}$CO($J = 1 - 0$) line (hereafter CO$_{10}$) with an effective critical density of $\sim 10^{2-3}$ cm$^{-3}$ and excitation temperature of $\sim 10$ K, thereby tracing the bulk of cold molecular gas.

In recent years, the $^{12}$CO($J = 2 - 1$) (hereafter CO$_{21}$) emission has emerged as the emission of choice. The higher frequency of this higher-J transition provides higher angular resolution and apparently better sensitivity to unit molecular mass. Large scale surveys such as Physics at High Angular resolution in Nearby GalaxieS (PHANGS; A. K. Leroy et al. 2021) utilize the CO$_{21}$ and convert it to molecular mass assuming a constant ratio between the two CO emission lines, $r_{21} (= \frac{I(\mathrm{CO}_{21})}{I(\mathrm{CO}_{10})})$. The motivation of this study is to reveal if there are spatial variations in $r_{21}$ and differences depending on galaxy morphology, which could have implications for studies estimating molecular mass assuming a constant ratio.

Theoretically, $r_{21}$ can change with gas volume density $n(\mathrm{H}_2)$, kinetic temperature $T_k$, and CO column density. Calculations based on the large velocity gradient (LVG) approximation (P. Goldreich & J. Kwan 1974; N. Z. Scoville & P. M. Solomon 1974) have shown that around the critical density of the two CO lines, the CO emission can be approximated as coming from the same region in a cloud (S. Sakamoto et al. 1997). Values in the range $r_{21} = 0.3 - 1$ are realized for gas with kinetic temperature $T_k > 10$ K and densities around and below $n(\mathrm{H}_2) \sim 10^3$ cm$^{-3}$ (S. Sakamoto et al. 1994). For dense ($n(\mathrm{H}_2) > 10^3$ cm$^{-3}$) and warm ($T_k > 50$ K) gas irradiated by UV photons, $r_{21} > 1$ may be realized (K. M. Gierens et al. 1992).

A number of observations have confirmed variations of $r_{21}$ within the Milky Way (MW). For individual clouds, S. Sakamoto et al. (1994) and T. Oka et al. (1998) show that the Orion A and B, and Taurus molecular clouds have significantly different averages of $r_{21} = 0.75 \pm 0.01$, $0.66 \pm 0.01$, and $0.53 \pm 0.01$, respectively. E. Falgarone et al. (1998) find that starless molecular clouds in the MW have a uniform $r_{21} = 0.65 \pm 0.15$. These indicate that $r_{21}$ can vary widely between clouds, and possibly that the presence of massive star formation may work to enhance $r_{21}$. Interestingly, the variation of $r_{21}$ within clouds is also similarly large, ranging from 0.5 to 1 within Orion. A. A. Nishimura et al. (2015) used high-resolution observations and point-by-point LVG calculations in Orion A to show that both the kinetic temperature and gas density (and correspondingly, $r_{21}$) have a gradient within the cloud, peaking at regions irradiated by a nearby OB association ($n(\mathrm{H}_2) \sim 2 \times 10^3$ cm$^{-3}$, $T_k \sim 100$ K, and $r_{21} \sim 1$), gradually decreasing toward more quiescent regions ($n(\mathrm{H}_2) < 10^3$ cm$^{-3}$, $T_k \sim 10$ K, and $r_{21} \sim 0.5$). Thus, density and temperature could both work to affect the actual $r_{21}$ in clouds.

At larger scales, a radial gradient in $r_{21}$ is found in the MW (T. Handa et al. 1993; S. Sakamoto et al. 1997; T. Sawada et al. 2001), decreasing from $\sim 1$ in the central kpc to $\sim 0.6$ at 8 kpc. The high value in the center is attributed to less opaque $CO_{10}$ with large velocity widths (T. Sawada et al. 2001), compared to clouds in the disk which are almost always optically thick. The physical origin of the radial gradient in $r_{21}$ is not clear, but observations of nearby galaxies is clearly more suited for this objective.

The measurements of $r_{21}$ in nearby galaxies have suffered substantially from observational uncertainties. The earliest studies by J. Braine & F. Combes (1992) and J. Braine et al. (1993) compared $r_{21}$ in the inner kiloparsec of 61 galaxies, finding an average of $r_{21} = 0.89$ but no systematic variations between barred and nonbarred, early and late types, or isolated and interacting galaxies. Although no formal uncertainties are presented in this work, these observations were single-point measurements of the centers conducted with different beamsizes for the two CO lines, so intrinsic differences in $r_{21}$ with respect to galaxy morphology or environment may have been washed out. L. P. Crosthwaite & J. L. Turner (2007) found variations of factor 6 in NGC 6946, ranging from $r_{21} = 0.35$ to $r_{21} = 2$ with half of the galactic disk having $r_{21} > 1$. However, the quoted flux uncertainties of 10 to 20% for each of the CO lines alone could account for bulk of this variation, illustrating how calibration of the telescope is crucial to the investigation of $r_{21}$. Spatial variations in $r_{21}$ could only be investigated in cases where data were taken under the best weather conditions and with special attention to calibration (J. Koda et al. 2012). K. Muraoka et al. (2016) find that $r_{21}$ is positively correlated with the kinetic temperature estimated from LVG modeling (F. F. S. van der Tak et al. 2007), which may indicate that $r_{21}$ is driven by star formation instead of the high-density gas prior to the onset of star formation.

Studies on a statistical sample of galaxies are still sparse, and consistently find elevated $r_{21}$ in the center but only weak radial trends (J. S. den Brok et al. 2021; Y. Yajima et al. 2021; A. K. Leroy et al. 2022). An important setback of these studies is that the two CO emission lines have been observed with a combination of single dish telescopes, whose systematic calibration uncertainties are still relatively large. For example, data used in Y. Yajima et al. (2021) combines $CO_{10}$ data obtained at the Nobeyama Radio Observatory (NRO) 45 m telescope (CO Multi-line Imaging of Nearby Galaxies (COMING): K. Sorai et al. (2019)) with calibration uncertainty estimates of 25%, and measurements at IRAM 30 m telescope, with uncertainty of 20%. Adding the random errors alone in quadrature, taking the ratio of two CO lines can already introduce random uncertainties of 30%. In addition, A. K. Leroy et al. (2022) show that even the global CO flux for a galaxy coming from different (or the same) telescopes can differ by $\sim 30$%. This converts to a range of 0.4 to 0.8 for an average of $r_{21} = 0.6$, thus potentially washing out (or creating) systematic trends in a galaxy. A recent work by R. P. Keenan et al. (2024b,2024a) found from global measurements of $> 100$ galaxies that $r_{21}$ correlates with the SFR and also on its offset from the SF main sequence. Although the uncertainties are relatively large ($\sim 20$%), the large sample size enables statistical verification of the correlation.

The calibration precision offered by the Atacama Large Millimeter/submillimeter Array (ALMA) largely surpasses most conventional radio telescopes. Its absolute flux calibration uncertainty (set by flux models of planets) is given as 5%. Absolute flux comparison with the Planck satellite (at the Band 3 frequency) indicates that the absolute uncertainty may be better, at 2.4% (G. S. Farren et al. 2021). Observing both CO lines with ALMA, we can thus expect to expose the detailed spatial variations in $r_{21}$.

J. Koda et al. (2020) use the Total Power (TP) array of ALMA, and found in M 83 that the ratio is high at the galactic center, decreases in the bar, increases at the bar end, and stays high in the subsequent spiral arms. At higher resolution ($\sim 200$pc), F. Egusa et al. (2022) find in NGC 1365 that the $r_{21}$ is lower within the bar and correlated with local star formation, but the ratio can change depending on the spectral components at a given position. Similarly, F. Maeda et al. (2022) find in NGC 1300 that $r_{21}$ is larger in giant molecular clouds associated with H$\alpha$ emission, and lower at larger (kpc) scales which they attribute to presence of extended gas at low volume density. J. S. den Brok et al. (2023) find that the bar regions in NGC 3627 and NGC 2903 have a systematically low $r_{21}$ compared to the center, with elevated values in the bar end.

In light of these studies of individual galaxies which suggest that the CO ratio can vary substantially depending on structure and characteristics within a galaxy (via a change in the local density or temperature, or column density), it is becoming necessary to obtain a more general view using a statistical sample of galaxies with varying global characteristics using accurately calibrated data sets. In this study, we will visit the question of whether there are large scale radial gradients in $r_{21}$, and if there are any differences depending on galaxy morphology.

## 2. ALMA-FACTS

The ALMA-Fundamental CO Transition Survey (FACTS) is a mapping survey of 12 nearby galaxies in the $CO_{10}$ line with the ALMA 12 m, 7 m and the TP arrays. The sample is a complete set of galaxies observed in PHANGS (A. K. Leroy et al. 2021), Spitzer SINGS (R. C. Kennicutt et al. 2003), and Herschel KINGFISH (R. C. Kennicutt et al. 2011), except that NGC 4569 is omitted due to its negative recession velocity outside the optimum frequency range of ALMA Band 3. The angular resolution and sensitivity are aimed to match that of ALMA-PHANGS observations. Details of the survey design, interferometric ALMA observations and reduction will be presented in Koda et al. (in prep., hereafter Paper I).

This paper presents an analysis of the TP observations obtained as part of ALMA-FACTS. These single dish observations provide zero-spacing information, so are free from uncertainties that can potentially be introduced in the imaging process of interferometer data. The TP-alone analysis presented here provides an important calibration of the value of $r_{21}$ averaged over $\sim 1/2$ of the galactic disk.

In Table 1, we list our samples from FACTS, their morphological classification, distance, and other geometric parameters. The optical classification of NGC 3521 (G. de Vaucouleurs et al. 1991) is SABbc, signifying the presence of a weak bar, but subsequent near-infrared studies (R. J. Buta et al. 2015; S. Dìaz-Garcìa et al. 2016) do not find evidence of a bar. We treat NGC 3521 as a nonbarred SA galaxy hereafter.

Table 1
Sample Properties of ALMA-FACTS

| Galaxy | Morphology | Distance (Mpc) | $D_{25}$ (arcmin) | $R_{A2}$ (arcmin) | Inclination (deg) | P.A. (deg) | $r_{21}$ (lum. weighted) | $r_{21}$ (disk+outlier) | $r_{21}$ (disk) | $r_{21}$ (outlier) |
|---|---|---|---|---|---|---|---|---|---|---|
| (1) | (2) | (3) | (4) | (5) | (6) | (7) | (8) | (9) | (10) | (11) |
| NGC 628 | SAc | 9.8 (1) | 10.5 | ⋯ | 19.8 | 20 | ⋯ | ⋯ | ⋯ | ⋯ |
| NGC 1097 | SBb | 15.4 (2) | 9.3 | 1.2 | 54.8 | 133.9 | 0.64 ± 0.02 | 0.65 ± 0.09 | 0.63 ± 0.07 | 0.71 ± 0.10 |
| NGC 1512 | SBa | 12.6 (2) | 8.9 | 0.79 | 68.3 | 65.2 | 0.52 ± 0.02 | 0.53 ± 0.10 | 0.51 ± 0.09 | 0.56 ± 0.11 |
| NGC 3351 | SBb | 9.33 (3) | 7.4 | 0.66 | 54.6 | 10.7 | 0.56 ± 0.02 | 0.58 ± 0.12 | 0.57 ± 0.12 | 0.66 ± 0.09 |
| NGC 3521 | SA* | 12.1 (2) | 11 | ⋯ | 60.0 | 162.8 | 0.61 ± 0.02 | 0.61 ± 0.05 | 0.61 ± 0.04 | 0.65 ± 0.09 |
| NGC 3621 | SAd | 6.55 (3) | 12.3 | ⋯ | 67.6 | 161.7 | 0.67 ± 0.02 | 0.66 ± 0.07 | 0.66 ± 0.07 | 0.67 ± 0.10 |
| NGC 3627 | SABb | 9.38 (3) | 9.1 | 0.94 | 67.5 | 168.1 | 0.61 ± 0.02 | 0.64 ± 0.06 | 0.63 ± 0.05 | 0.69 ± 0.06 |
| NGC 4254 | SAc | 13.9 (4) | 5.4 | ⋯ | 20.1 | 66 | 0.63 ± 0.02 | 0.65 ± 0.07 | 0.63 ± 0.06 | 0.74 ± 0.09 |
| NGC 4321 | SABbc | 14.3 (2) | 7.4 | 1.0 | 24.0 | 158 | 0.55 ± 0.02 | 0.59 ± 0.08 | 0.58 ± 0.06 | 0.66 ± 0.12 |
| NGC 4536 | SABbc | 14.5 (2) | 7.6 | 1.1 | 73.1 | 120.7 | 0.69 ± 0.02 | 0.70 ± 0.06 | 0.67 ± 0.06 | 0.72 ± 0.05 |
| NGC 4579 | SABb | 16.4 (4) | 5.9 | 0.54 | 41.9 | 90.2 | 0.56 ± 0.02 | 0.58 ± 0.10 | 0.57 ± 0.10 | 0.67 ± 0.12 |
| NGC 4826 | SAab | 4.41 (5) | 10 | ⋯ | 64.0 | 114.0 | 0.69 ± 0.02 | 0.68 ± 0.06 | 0.67 ± 0.05 | 0.74 ± 0.07 |

**Note.** (2) Galaxy morphology taken from RC 3 (G. de Vaucouleurs et al. 1991). *For NGC 3521, G. de Vaucouleurs et al. (1991) lists this as a SABbc galaxy, but we treat is as a nonbarred SA for reasons explained in Section 2. (3) Redshift-independent distance, numbers in parentheses are the references. In order of preference, measurements taken from Cepheid variables, tip of RGB, and Tully–Fisher relation. References : 1. K. B. W. McQuinn et al. (2017), 2. R. B. Tully et al. (2016), 3. W. L. Freedman et al. (2001), 4. R. B. Tully et al. (2013), 5. G. S. Anand et al. (2021). (4) Isophotal diameter at the major axis, taken from RC 3 (G. de Vaucouleurs et al. 1991). (5) Characteristic radius of the bar, measured as the position at the maximum amplitude of the $m = 2$ Fourier component (S. Dìaz-Garcìa et al. 2016). (6) (7) Inclination and position angle taken from HyperLEDA (D. Makarov et al. 2014) when available, or from N. Kuno et al. (2007) and K. Sorai et al. (2019). The values here were used as initial values for the tilted-ring model fitting in Section 4.1. For more recent values, see P. Lang et al. (2020). (8) Luminosity-weighted average, measured as in Section 3.1. Errors are assumed to be 3%, assuming an absolute flux calibration uncertainty of 2.4% for both CO lines and adding them in quadrature. (9) Average CO line ratio $r_{21}$ within the area used for the PV diagram along the major axis. Errors are the standard deviation of the values. (10) Same as (9) in the disk component. (11) Same as (9) in the outlier component. See Section 4.1.

Table 2
ALMA-FACTS TP Observation Parameters

| Galaxy | Observation | Map Size (arcmin) | Map P.A. (deg) | $T_{sys}$ (K) | $N_{EB}$ | Scaling Factor (Std. Dev., %) | R.m.s. (mK) |
|---|---|---|---|---|---|---|---|
| (1) | (2) | (3) | (4) | (5) | (6) | (7) | (8) |
| NGC 628 | 2022 Oct 22 – 2023 Mar 14 | 5.5 × 4.4 | 135 | 110–144 | 35 | 2.3 | 0.96 |
| NGC 1097 | 2023 Jan 1 – 2023 Mar 7 | 6.0 × 3.8 | 145 | 95–128 | 23 | 2.2 | 0.90 |
| NGC 1512 | 2023 Mar 25 – 2023 Apr 9 | 4.3 × 3.5 | 36.8 | 105–146 | 15 | 5.9 | 1.1 |
| NGC 3351 | 2023 Jan 21 – 2023 Feb 20 | 3.9 × 3.9 | 0 | 117–150 | 17 | 3.7 | 1.2 |
| NGC 3521 | 2023 Mar 2 – 2023 Mar 14 | 6.3 × 3.5 | 160 | 99–120 | 18 | 2.1 | 1.0 |
| NGC 3621 | 2023 Mar 18 – 2023 Apr 1 | 5.7 × 3.8 | 160 | 102–131 | 16 | 2.2 | 1.1 |
| NGC 3627 | 2023 Apr 6 – 2023 Apr 17 | 5.7 × 3.8 | 0 | 105–142 | 23 | 1.0 | 1.0 |
| NGC 4254 | 2023 Apr 29 – 2023 May 08 | 5.1 × 4.6 | 0 | 90–128 | 18 | 1.5 | 0.95 |
| NGC 4321 | 2022 Dec 29 – 2023 Mar 4 | 5.1 × 4.8 | 90 | 95–132 | 29 | 2.6 | 0.97 |
| NGC 4536 | 2023 Apr 16 – 2023 Apr 21 | 5.7 × 3.3 | 119 | 86–105 | 11 | 2.6 | 1.0 |
| NGC 4579 | 2023 Mar 12 – 2023 Apr 9 | 4.3 × 3.5 | 52 | 102–122 | 8 | 2.4 | 1.4 |
| NGC 4826 | 2023 Mar 4 – 2023 Mar 22 | 4.1 × 3.1 | 112 | 120–150 | 15 | 3.0 | 1.4 |

**Note.** (3) Size of mapped rectangular area in arcminutes. (4) Position angle of the long side of the map, measured from north counterclockwise. (5) 10–90 percentile range of the system temperature during the observations. (6) Number of execution blocks. (7) Standard deviation of the scaling factors between execution block pairs, see Section 2.2. (8) Map r.m.s. noise measured in 10km/s channel.

### 2.1. Observation

The observations were carried out using the TP Array, a part of the Atacama Compact Array (ACA; S. Iguchi et al. (2009)), in 2022 Oct–2023 May. The array consisted of 2–4 (3.33 on average) 12 m antennas and the ACA Correlator (T. Kamazaki et al. 2012). A 500-MHz wide, 2048-channel, dual-polarization spectral window (282 kHz resolution, i.e., 0.75 km s$^{-1}$ at 115 GHz) was placed at the frequency of the CO$_{10}$ line in one of the four basebands. Three more spectral windows were set up to cover the CN lines, C$^{17}$O($J = 1 - 0$) line, and radio continuum emission but are not reported in this paper. The half-power beamwidth (HPBW) at 115 GHz was ≈50″.6.

A rectangular field for each galaxy was mapped using the on-the-fly observing technique, along the long side of the field. A line-free reference position was visited before and after every raster row. The mapped area is larger than the area mapped with the TP array in CO$_{21}$ by PHANGS. The map size, orientation, and other observing parameters are listed in Table 2.

### 2.2. Data Reduction

We reduced both CO$_{10}$ from our FACTS survey and CO$_{21}$ as described in J. Koda et al. (2020, 2023), which is briefly summarized below. For the CO$_{21}$, we used archival

data (2012.1.00650.S, 2015.1.00925.S, 2017.1.00886.S, 2018.1.01651.S) that were taken as parts of the PHANGS (A. K. Leroy et al. 2021) survey. Only the calibrated TP array data cube was used.

The data were calibrated and mapped using the Common Astronomy Software Applications (CASA; CASA Team et al. (2022)). The data were calibrated into the antenna temperature ($T_A^*$) scale with the chopper-wheel method. The residual features caused by the atmospheric $O_3$ lines were removed by applying the method of T. Sawada et al. (2021). Linear spectral baselines were subtracted.

The calibrated data were mapped onto a regular grid of HPBW/9 ($\approx 5''.62$ for $CO_{10}$, $\approx 2''.81$ for $CO_{21}$) using the prolate spheroidal function with a support radius of 6 pixels (F. R. Schwab 1980, 1984) following the standard imaging procedure for TP Array. The effective spatial resolution after gridding convolution is $\approx 56''.6$.

We generated separate data cubes for all execution blocks (EBs). Some of the cubes showed significantly poorer signal-to-noise (S/N) ratios than the others due to bad weather and were removed. We calculated the relative flux scales among the cubes, by calculating the flux ratios and errors of all of their pairs, and solved for the scales by inverting the design matrix. The median of the relative flux scaling coefficients is normalized to unity. This method does not use any particular EB as a reference and distributes errors among the scaling coefficients of all EBs. Hence, it does not suffer from a potential anomaly in a reference EB. The standard deviation of the normalized coefficients is given in Table 2. The derived scaling factors were applied to EBs to coadd all the data into one final cube for each galaxy.

The $CO_{21}$ image was then smoothed to the $CO_{10}$ beam size $56''.6$ using a Gaussian kernel in the CASA task *imsmooth*.

For the purpose of this paper, the velocity was binned to $10 \mathrm{km\,s^{-1}}$ for both lines.

The integrated intensity (moment 0) map of the $CO_{10}$ line is shown in panel b) of Figure 1, created using only channels where CO was detected. The r.m.s. noise level of the data cube was measured over the map area in the velocity range of $100 \mathrm{km\,s^{-1}}$ width before and after the channels where CO was detected, and is tabulated in Table 2. The noise level of the integrated intensity map is estimated by multiplying the cube r.m.s. with the square root of the full velocity width of the channels with CO detection. The intensity weighted velocity (moment 1) map is shown in panel (c), which was created using regions detected at over $5\sigma$ in the integrated intensity map.

For the mean distance of our sample galaxies 11.2 Mpc, our spatial resolution of $56''.6$ corresponds to a projected scale of 3.1 kpc.

We exclude NGC 628 from this paper because the very small velocity width ($\sim 70 \mathrm{km/s}$) of this galaxy severely limited our analyses using position–velocity (PV) diagrams in Section 3.3.

## 3. Results

### 3.1. Luminosity-weighted CO Ratio

We first estimate the ratio $r_{21}$ derived as a ratio between the total flux of CO within the images. The total CO flux for both lines was measured within the whole field of view of the integrated intensity map of $CO_{21}$. The values are listed in Table 1. $CO_{10}$ was distributed well outside of the area mapped in $CO_{21}$, so the luminosity-weighted $r_{21}$ does not completely represent the true value expected from global flux. However, they can be compared directly to values in the literature that do not spatially resolve the galaxies. The average value of luminosity-weighted mean is $r_{21} = 0.61$ with a standard deviation of 0.06, consistent with previous studies (Y. Yajima et al. 2021; J. S. den Brok et al. 2021; A. K. Leroy et al. 2022).

### 3.2. Ratio Map

The map of the CO line ratio $r_{21}$ derived from the integrated intensity maps of the two CO lines is shown in panel (d) of Figure 1. Note that $r_{21}$ is shown only for pixels where both CO lines are above an S/N of 5. The mean S/N in the data ranged from 12 in NGC 1512 to 58 in NGC 3627. The map-based cut of S/N > 5 is used only for presenting the map, but for the analysis presented in this paper after Section 3.3 we further clipped data at S/N of 10 or above.

Although the spatial resolution is coarse ($\sim 3 \mathrm{kpc}$), variation of $r_{21}$ is already evident at some level. The central regions in NGC 3351, NGC 4254, NGC 4321, and NGC 4579 have elevated $r_{21}$ compared to the disks. NGC 3627 shows higher $r_{21}$ in the spiral arm. A notable feature in NGC 1097, NGC 4536, and NGC 4826 are peaks in the direction of the minor axis, which can be explained by the outlier components discussed in Sections 3.4 and 3.5. It should be noted that the variations we find in this study using the TP array are the minimal variations within the galaxy; larger variations at smaller spatial scales (e.g., with individual star-forming regions) are smeared out. In spite of the spatial variation in $r_{21}$, the low angular resolution of the TP array does not allow for a detailed study, in particular with respect to large scale kinematic signatures over the galaxy.

### 3.3. PV Diagram

We exploit the high S/N afforded by ALMA to spectrally resolve the variation in $r_{21}$, and investigate its relation to kinematic signatures. We use the PV diagram along the major axes of each sample.

The left-hand panels in Figure 2 show the distribution of $r_{21}$ in the PV diagram along the major axes for our ALMA FACTS sample. The PV diagrams for both $CO_{10}$ and $CO_{21}$ data are created using CASA task *impv*, using the position angles presented in Table 1. The widths of the slits are set to $1'$, comparable to the beam size of the TP array data. The lengths of the slits were determined to maximize the length within the $CO_{10}$ map. Pixels within half a beam ($28''$) from the edges of the maps are shown but were excluded from further analysis. The r.m.s. noise in the $CO_{10}$ PV diagram, $\sigma$, was measured in emission free regions. The $CO_{21}$ PV diagram was then divided by the $CO_{10}$ PV diagram for regions over $20\sigma$ to obtain the distribution of $r_{21}$. For NGC 1512 and NGC 4579, a threshold of $10\sigma$ was used because the noise levels for these two galaxies were relatively high. This sigma cut corresponds to regions with S/N over 10 for a dominant part of the $CO_{21}$ PV diagram. Considering that the typical range of $r_{21}$ is 0.4 to 0.8 within and between galaxies, our high noise level cuts enable us to detect small variations in $r_{21}$ of $\lesssim 0.05$.

Column (9) in Table 1 gives the average $r_{21}$ calculated from the PV diagram. The given error indicates the intrinsic scatter of the $r_{21}$. In contrast to the luminosity-weighted $r_{21}$ calculated in Section 3.1 over the whole map, the values here are restricted to pixels within the area of the PV diagram and weighted with respect to area in PV space. The two estimates therefore do not necessarily have to match, but they are consistent and well within the standard deviation. This indicates that our estimate of $r_{21}$ along the major axis is representative of the whole galaxy.

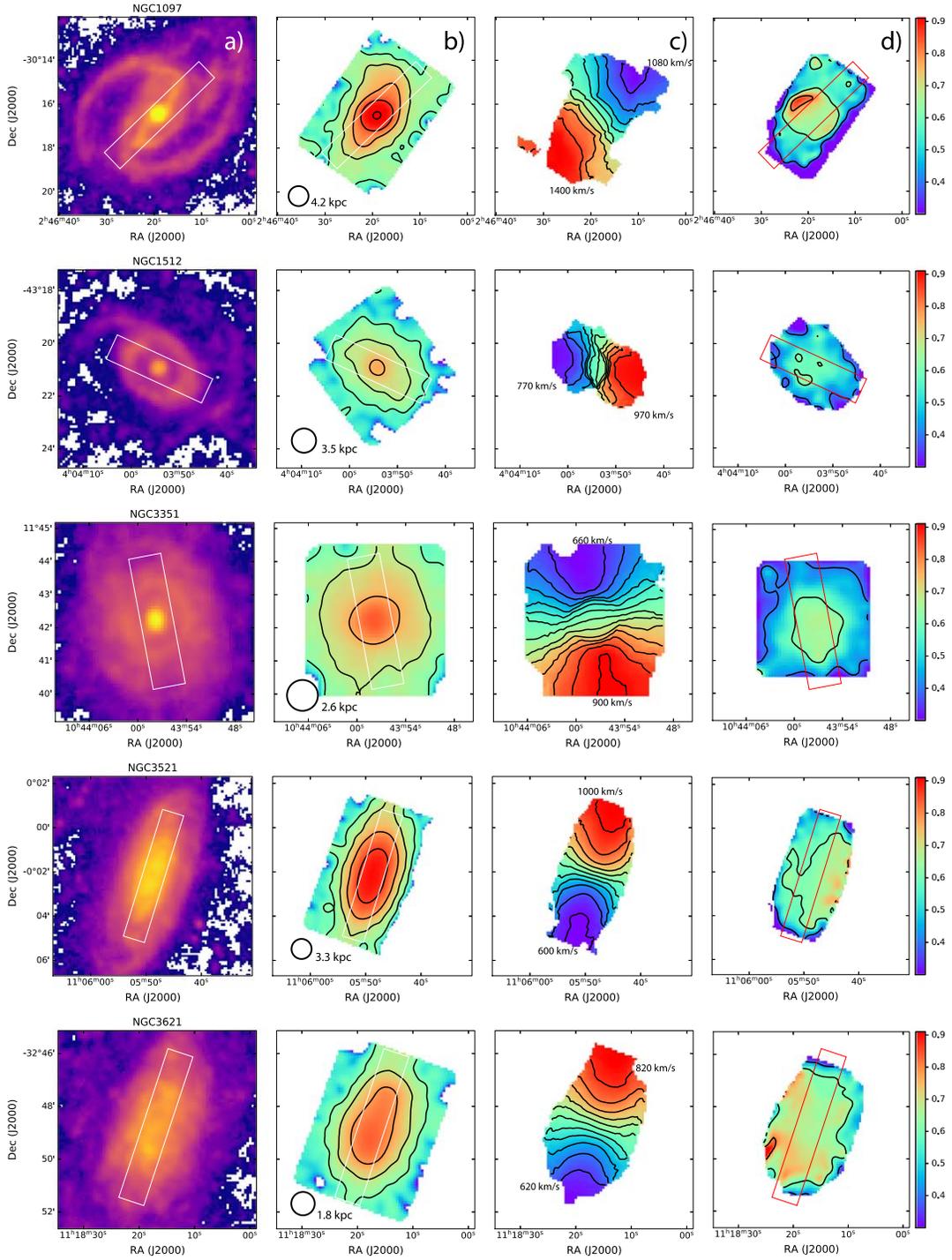

**Figure 1.** Panel (a) : Herschel SPIRE 250 $\mu$m image from KINGFISH (R. C. Kennicutt et al. 2011). The rectangles here and in panels (b) and (d) indicate the region where the PV diagram was created. (b) : Integrated intensity (moment 0) map of $^{12}$CO($J = 1 - 0$) from the ALMA TP array. The black circle represents the beam FWHM of 56$\farcs$6, along with its projected physical scale. Contours are drawn from 5 times the r.m.s. noise level, in increments of factor 3. (c) Intensity weighted velocity (moment 1) map. Contours are drawn in steps of 20, 30, or 40 km s$^{-1}$. The maximum and minimum velocities of the drawn contours is indicated explicitly. (d) $r_{21}$ map, shown for the area where CO$_{21}$ map overlap. Contours are drawn at 0.4, 0.6, and 0.8.

### 3.4. Spatial and Spectral Variation of $r_{21}$

Clear spatial and spectral variations in $r_{21}$ are apparent from the PV diagrams. While all galaxies show evidence of a rotating gas disk with relatively low $r_{21}$ around 0.5 to 0.7, most galaxies show elevated $r_{21}$ of around 0.8 at edges of the PV diagrams. The increase in $r_{21}$ at these edges of the PV diagram cannot be explained by effects coming from low S/N because our sigma cut of 20 times the $CO_{10}$ r.m.s. ensures that random variations in $r_{21}$ are minimized. Considering that our angular resolution is lower than the typical spatial extent of the edges ($\sim 10 - 20''$ for a given spectral channel), the higher $r_{21}$ at the edges should not be interpreted as a spatial variation at a given velocity channel but rather as increased $r_{21}$ at the low- and/or high-velocity components of a spectrum at a given position.

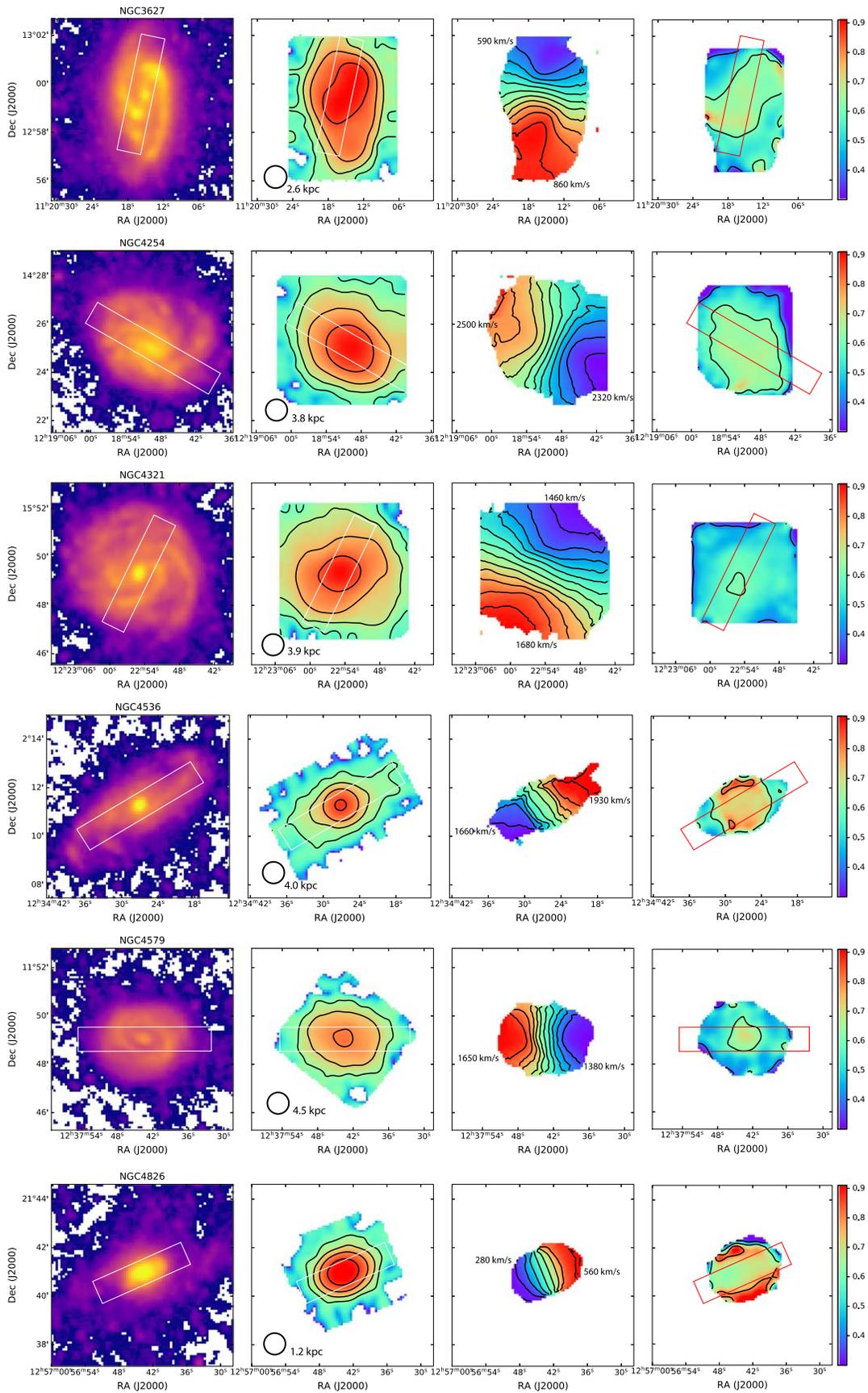

**Figure 1.** (Continued.)

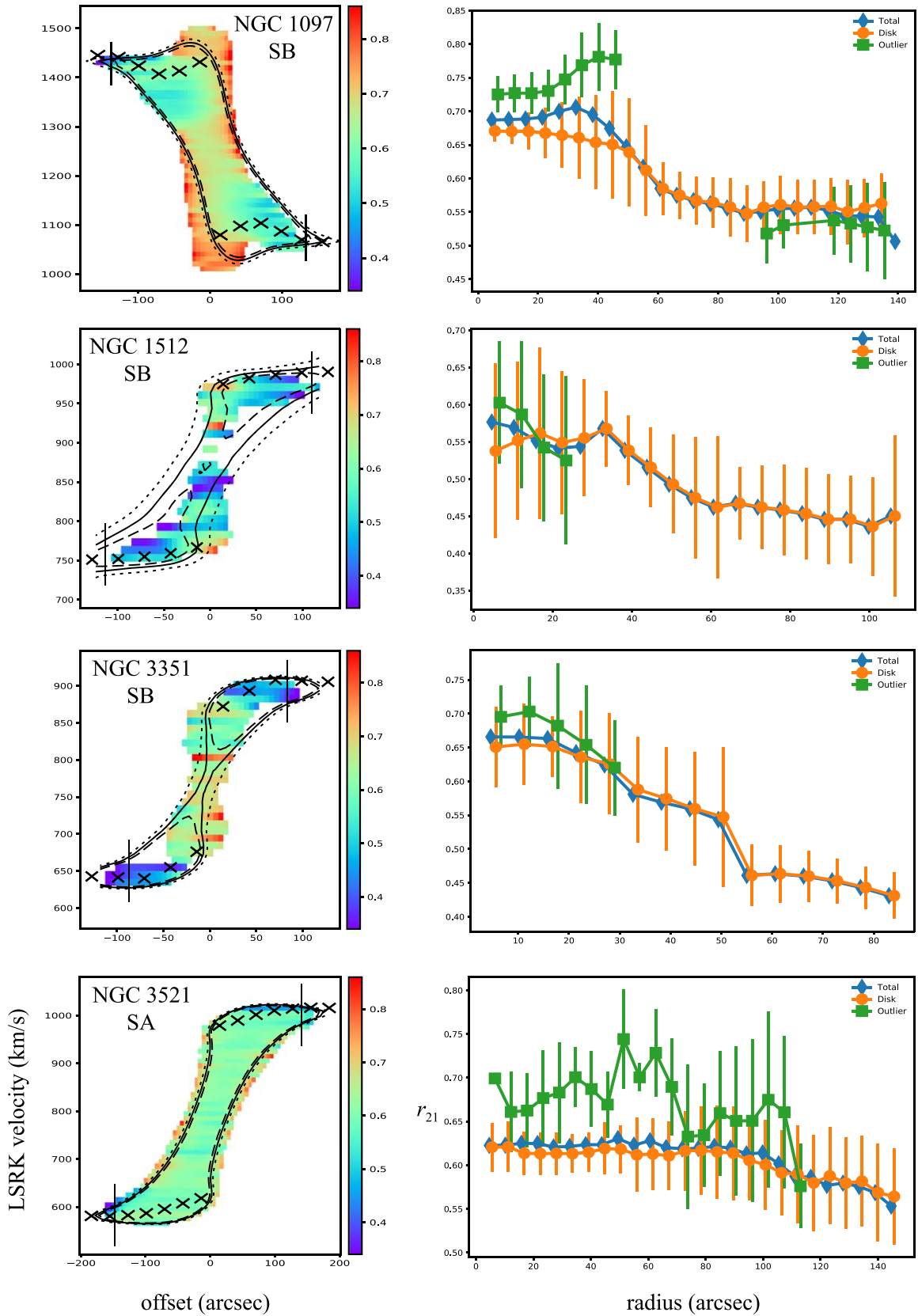

**Figure 2.** Left-hand panels: PV diagram of $r_{21}$ along the major axis shown in panel (a) of Figure 1. Solid contours and crosses correspond to the region modeled as a rotating disk and the projected rotation velocity, respectively, as explained in Section 4.1. The dashed and dotted lines correspond to the model PV contours $5\sigma$ above and below that indicated by the solid line. The short vertical lines indicate the radii out to which the radial plots in the right panel are calculated, and are limited by the size of the $CO_{21}$ image. Right-hand panels: Radial change of $r_{21}$ calculated from the left-hand panel. Filled orange circles are for the disk region enclosed by solid contours on the left-hand panel, and green squares are the outliers. Blue diamonds are for disk and outlier combined. Only radii with 3 or more data points are plotted. Errorbars represent the standard deviation of the values on the PV diagram.

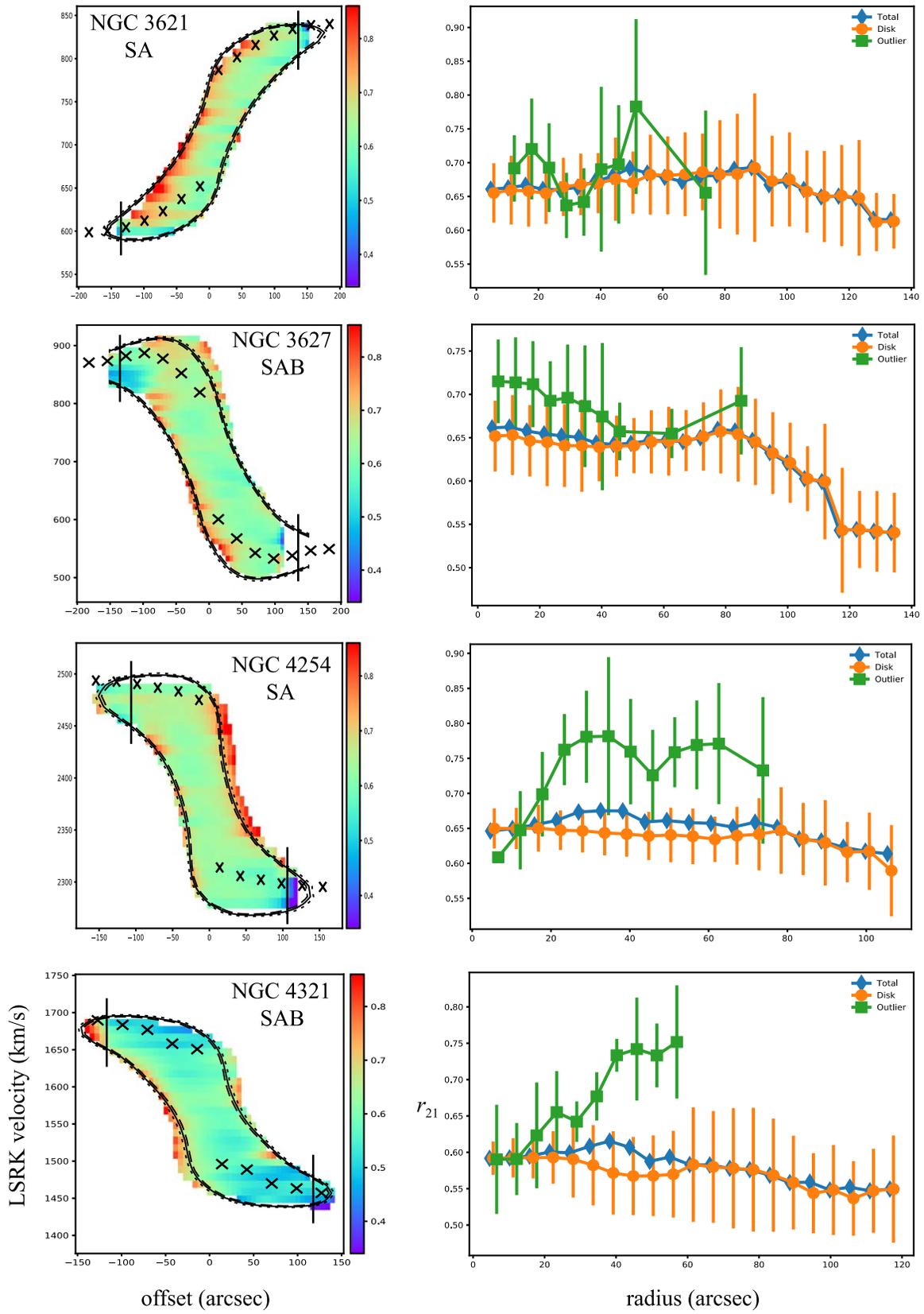

**Figure 2.** (Continued.)

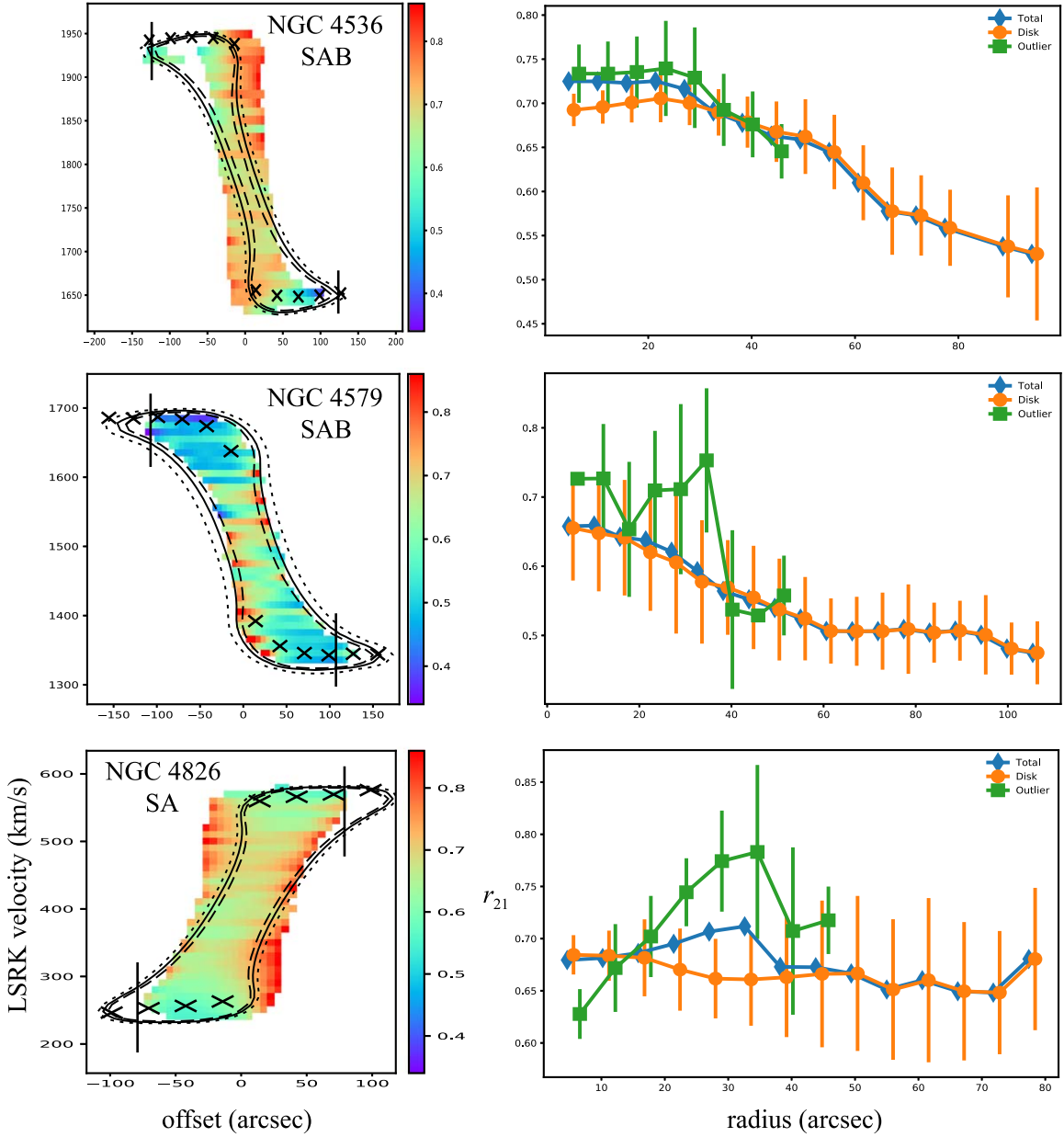

**Figure 2.** (Continued.)

### 3.5. Physical Origin of High $r_{21}$ at Edges of the PV Diagram

Variations in $r_{21}$ with respect to velocity (at a given position) have been pointed out by F. Egusa et al. (2022) in NGC 1356, who argue that different ratios between spectral components may correspond to different kinematical signatures, although their data suggest that their nondisk component (outflow) has a low $r_{21}$, contrary to our edges where the ratio is high.

The elevated $r_{21}$ in Figure 2 is found preferentially at the forbidden velocities (i.e., at the opposite side of the main body of galactic rotation). This may be explained by an unresolved compact source with large $r_{21}$ in the nuclei, which will be observed in the PV diagram as nearly vertical structures with width of $\sim 5.6''$. If the CO flux of the disk is dominant over the compact component at larger radii, effects of large $r_{21}$ will not show in the positions of the rotating disk but will be observable in the forbidden velocities where the disk component becomes negligible. NGC 1097, NGC 3351, NGC 4536, and NGC 4826 are examples of such PV diagrams with clear signatures of unresolved compact sources. The physical origin of these sources can be associated with dense circumnuclear gas, its associated star formation, and/or AGN activity, which can increase the gas volume density, kinetic temperature, or alter abundances and opacity to give higher $r_{21}$. Another potential mechanism for creating large $r_{21}$ in the forbidden velocity (i.e., counter-rotating side) could be due to components of the gas that traveled toward the center along the $x_1$ orbit in a nonaxisymmetric potential (i.e., a bar inflow), which then move to the the other side of the disk instead of settling in the $x_2$ orbit (so called overshooting) (M. W. Regan et al. 1999; E. Athanassoula & M. Bureau 1999; J. Koda & K. Wada 2002; H. P. Hatchfield et al. 2021; M. C. Sormani et al. 2023). Such overshooting gas will eventually converge on the dust lane on the other side of the disk, where it can temporarily move outward against the inflow (R. G. Tress et al. 2020), placing the gas at a forbidden velocity. There the increased local gas density alone may increase $r_{21}$, or further by the star formation which may be induced. The spatial resolution of our TP array does not allow for detailed investigation on what process is responsible for the elevated $r_{21}$ at the edges of our PV diagram, and these will be treated in forthcoming papers using high spatial resolution data sets.

## 4. Discussion

### 4.1. Modeling the Galactic Disk

One of our objectives in this paper is to investigate the radial trend of $r_{21}$ in our galaxies. The high $r_{21}$ in the PV diagram edges could be due to spatially unresolved structures near the nuclei (as discussed in the previous section), and could work as contaminants in the overall trend in the disk.

In order to distinguish between the edges of the PV diagrams and the normal galactic rotation in a physically motivated manner, we use 3DBarolo (E. M. Di Teodoro & F. Fraternali 2015) to fit the data cube with a tilted-ring model (D. H. Rogstad et al. 1974). The $CO_{10}$ data cube was modeled using the position angle and inclination given in Table 1 as initial parameters, with concentric rings 28″8 in width. The position angle, inclination, and rotation velocity were allowed to change in each ring. The radial gas inflow/outflow was fixed to 0. Meanwhile, 3DBarolo outputs include the best-fit $CO_{10}$ rotation curve, radial CO profiles, and the corresponding model cube.

After the model fit, a PV diagram along the major axis was created with the modeled CO cube using the same parameters as done for deriving the observed $r_{21}$ ratio. We use this model $CO_{10}$ PV diagram, indicating pure-circular rotation, to distinguish rotating gas with gas in noncircular rotation. A contour in the model PV diagram was drawn at $30\sigma$ (i.e., $10\sigma$ above the threshold level used in calculating the observed ratio), inside which the observed data were assigned to be the "disk" component. The gas components filtered out from the disk component (i.e., on and outside the designated contour in the model PV diagram) are hereafter labeled as "outlier" components. For NGC 1512 and NGC 4579, a $10\sigma$ threshold was used for defining the contour of the disk component, for the same reason as explained in Section 3.3. The threshold level designated here filters out gas velocities that deviate from a pure rotating disk. This choice does not mean that the components within the disk contour all follows a pure-circular rotation; instead, it singles out the component that clearly deviates from the circular rotation at any orientation and projection.

The S/N cuts used above are somewhat arbitrary. They are chosen to be a reasonable balance between distinguishing the high ratio at the edges, while being applicable uniformly throughout most of the samples. In Figure 2 we also show the

**Table 3**
$r_{21}$ Measured in the PV Diagram

|  | Disk + Outlier | Disk | Outlier |
|---|---|---|---|
| All galaxies | 0.63 ± 0.09 | 0.62 ± 0.08 | 0.69 ± 0.10 |
|  | (0.63 ± 0.09) | (0.61 ± 0.09) | (0.69 ± 0.10) |
| SA | 0.65 ± 0.07 | 0.64 ± 0.06 | 0.70 ± 0.10 |
|  | (0.65 ± 0.09) | (0.64 ± 0.06) | (0.70 ± 0.09) |
| SAB | 0.63 ± 0.09 | 0.61 ± 0.08 | 0.70 ± 0.08 |
|  | (0.63 ± 0.09) | (0.61 ± 0.08) | (0.70 ± 0.08) |
| SB | 0.62 ± 0.11 | 0.60 ± 0.10 | 0.68 ± 0.11 |
|  | (0.59 ± 0.11) | (0.57 ± 0.11) | (0.65 ± 0.12) |

**Note.** Mean and standard deviation of $r_{21}$ for each galaxy category and region. The lower row for each category lists the value for a normalized distribution, explained in Section 4.2.

contour level $5\sigma$ above and below the threshold explained above to illustrate how our choice of the sigma threshold affects the results. For most galaxies, the contour levels are sufficiently close, indicating that the distinction of disk and outliers are robust against our choice of threshold. For NGC 4579, changing the threshold to a lower value ($5\sigma$ instead of $10\sigma$) results in the whole PV diagram being defined as a disk; the increase in $r_{21}$ at the edges of the PV diagram is not prominent in this galaxy, and should be interpreted with care.

We used $CO_{10}$ to model the galaxy instead of $CO_{21}$ because we can expect its lower critical density to allow us to trace the bulk of gas and also because the area mapped in $CO_{10}$ is wider compared to $CO_{21}$, thereby giving a better representation of the whole galaxy. The PV diagrams of both CO lines were nearly identical, and using $CO_{21}$ to model the galaxy parameters produced consistent results in the analyses, even though we use the S/N ratio for the distinction of disks and outliers. This comes from the survey design of FACTS where the sensitivities of the observations were designed to be consistent with the $CO_{21}$ observations (see Paper I).

For NGC 1097, NGC 3351, NGC 4536, and NGC 4826, the PV diagrams clearly show components which clearly deviate out from the circular rotation model, with apparently high $r_{21}$. This suggests that for these sources, the high $r_{21}$ is related to noncircular motions, galactic center activities, or other events that deviate from the pure-circular motions. For other galaxies such as NGC 3521 and NGC 4579, the outliers may at least partly be explained by our choice of the sigma threshold because expanding the contour may enclose the outlier component.

The left-hand panels in Figure 2 show the observed PV diagram along with the contour outlining the disk component and the rotation curve to illustrate the gas motion in the disk.

### 4.2. Number Distribution of $r_{21}$

The average value of $r_{21}$ for each kinematic component (disk and outlier) in each galaxy are tabulated in Table 1. For all galaxies, $r_{21}$ is larger for the outlier components compared to the disk. Table 3 gives the mean and standard deviation for disks, outliers, and both combined for different galaxy morphology. The average value of $r_{21}$ for all our galaxies (disk and outlier combined) is $r_{21} = 0.63 \pm 0.09$, which is consistent with the average of the luminosity-weighted mean (0.61 ± 0.06) in Section 3.1. For the comparison among our sample galaxies, the systematic uncertainty that affects all galaxies equally should cancel out in a relative sense. For an average of 0.63 and a flux calibration uncertainty of 2.4%, the random error in $r_{21}$ propagates to $\delta r_{21} = 0.021$. The standard deviation is always larger than $\delta r_{21}$, so the spread among galaxies and between galaxy categories can be regarded as real. In panel (a) of Figure 3 we show the number distribution histogram of the $r_{21}$ values in the disk and outliers for all of the galaxies combined. The bins in the histogram have width of 0.027, i.e., larger than $\delta r_{21}$. Since the number of pixels in the PV diagram differ between galaxies, we also show the same distribution but normalized by the number of data points for each galaxy, which are then coadded and divided by the number of sample galaxies. This normalized distribution is shown in blue circles and orange squares. The normalized distribution matches the non-normalized distribution very well, and a Welch's t-test between the disk and outliers in the normalized data set gives a p-value of 0.009, indicating that the disk and outliers likely come from different distributions. When the t-test is performed between disk and outliers separately for SA, SAB, and SB galaxies, the p-values are 0.024, 0.007, and 0.002, respectively. These indicate that the disk and outliers have a distinctive distribution regardless of morphology. The $r_{21}$ distribution for each morphology category is shown in panels (b) to (d). The outliers have a mean $r_{21}$ that is higher by nearly 0.1 compared to the disk. The contribution of outlier pixels to the global distribution is small, however, and the distribution of $r_{21}$ is dominated by pixels attributed to the disk. The higher end of the distribution for the disk is contributed to a large degree by pixels in the PV diagram that are adjacent to the outliers, and are likely due to our somewhat arbitrary distinction between the kinematic components. The lower end of the disk distribution is dominated by regions at large galactic radii in the disk.

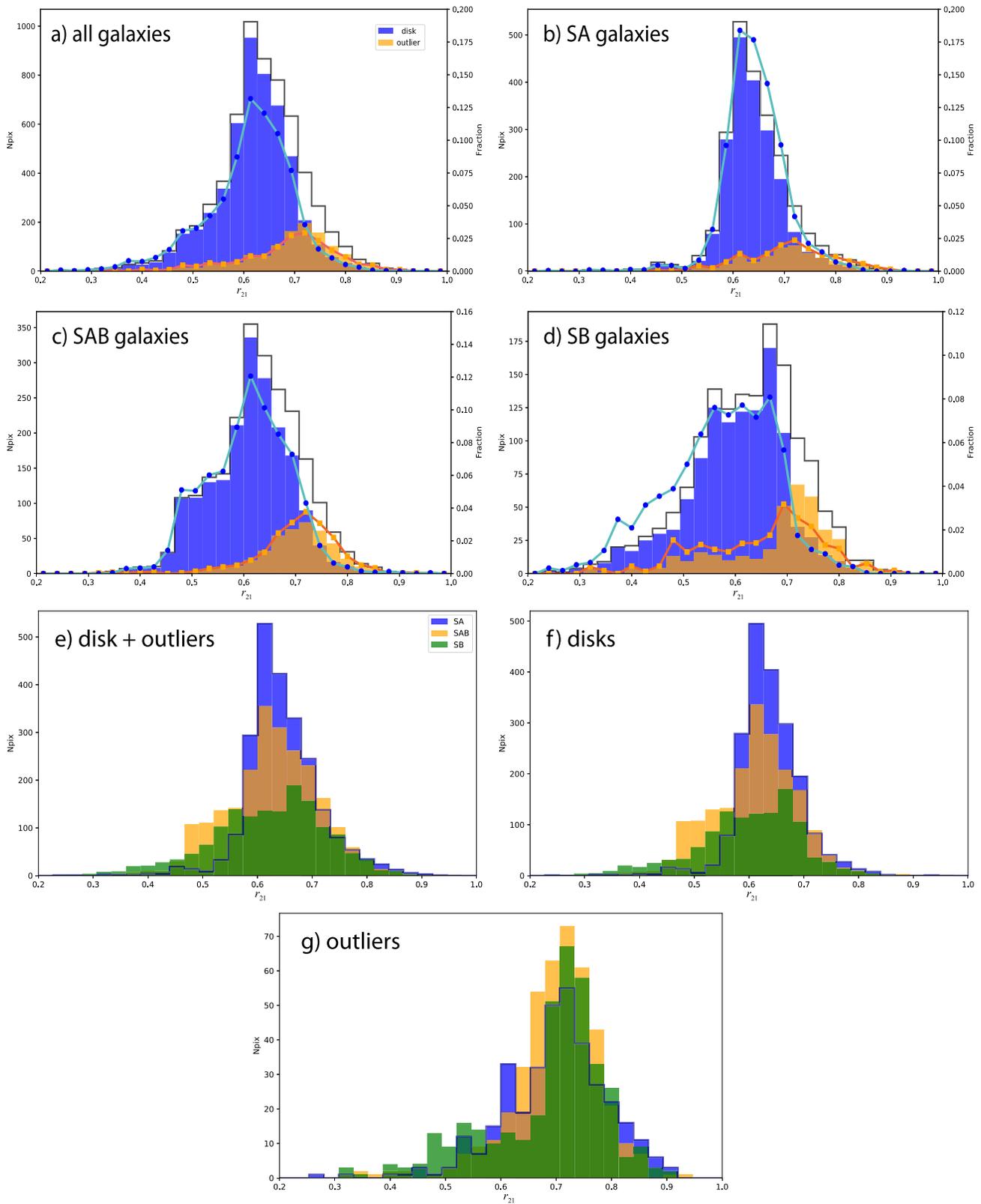

**Figure 3.** Number distribution of $r_{21}$ in the PV diagram pixels. Panel (a): For all galaxies, between disk (blue), outliers (orange), and both combined (black lines). The light blue and dark orange lines correspond to the normalized distribution, where data points from individual galaxies are normalized with respect to their number of data points, coadded, and then divided by the number of sample galaxies. The $y$-axis on the right-hand side indicates the fraction in the normalized distribution. Panels (b), (c), (d): Same for (a) but for SA, SAB, and SB galaxies only, respectively. Panel (e): Between SA, SAB, and SB galaxies with the disk and outliers combined. The distribution for the SA galaxies is outlined to aid the view. Panels (f) and (g), same as panel (e) but for disk and outliers, respectively.

In panels (e), (f), and (g), we compare the distribution between different morphology separately for disk and outlier combined, disk, and outliers. A notable feature is that the number of data points in the disk (panel (f)) are significantly smaller in SB compared to SA galaxies, but comparable in the outliers (panel (g)), indicating that barred galaxies harbor a larger fraction of outliers compared to nonbarred galaxies. This mirrors the fact that SB galaxies cannot be modeled by simple tilted-ring models with no assumed radial motion, which coincidentally also have high $r_{21}$. They are likely to be components distributed as unresolved nuclear sources proposed in Section 3.5. The mean and standard deviation of $r_{21}$ values in Table 3 indicate that there are no significant differences between galaxies with and without bars. A Welch's t-test was performed between SA, SAB, and SB galaxies separately for the normalized distribution of disks and outliers, and none indicated any significant differences in the distribution. High-resolution studies of the strongly barred galaxy NGC 1300 (F. Maeda et al. 2020, 2022) have found low $r_{21}$ values of 0.2 to 0.4 in the bar, which they attribute to extended (and presumably low density) gas missed by interferometers. F. Egusa et al. (2022) derive a high-resolution $r_{21}$ map for a strongly barred galaxy, NGC 1365, which also seems to indicate a lower $r_{21}$ of $\sim 0.5$ in the large scale bar compared to the spiral arm. Our $r_{21}$ maps and PV diagrams cover the large-scale stellar bars for the barred samples. For NGC 1097 and NGC 1512 (both SB), the direction of the bar roughly coincides with the galaxy major axes, thus the PV diagram encompasses the bulk of the bar. For NGC 3351 (SB), the bar is almost perpendicular to the major axes, but the bar's length is short ($0\farcm66$), so that the PV diagram still includes gas in the bar. As we show in Section 4.3, we do not see any decrease in $r_{21}$ within the bar or a rise in $r_{21}$ at radii corresponding to bar ends. The fact that we do not see lower $r_{21}$ for barred galaxies may simply reflect our low spatial resolution.

### 4.3. Radial Gradient

The radial profiles of the disk, outlier, and the two components combined are shown in the right-hand panel of Figure 2. The width of the radial bins is $5\farcs6$, corresponding to the pixel size, i.e., 1/10 of the beam. Thus, the adjacent radial bins are not completely independent. Since $r_{21}$ is an intensive quantity (i.e., independent of mass), its profile for the outlier can be traced toward the center, even though the amount of gas sampled there is very small. For many of the galaxies, the outliers peak not at the center but around a radius of $30''-40''$. This is explained by the raised $r_{21}$ at the edges of the PV diagram, especially from the forbidden velocities, detailed in Section 3.4. This is the cause of $r_{21}$ peaks in the direction of the minor axis in the $r_{21}$ map (panel (d) in Figure 1), in galaxies NGC 1097, NGC 4536, and NGC 4826. These galaxies have prominent outlier components.

The low angular resolution of our data does not allow us to determine the mechanism causing the outlier components (although possibilities are discussed in Section 3.5), we will focus on the radial trends of the disk component.

While the disk of some galaxies shows a clear radially decreasing trend in $r_{21}$ (i.e., NGC 1097, NGC 1512, NGC 3351, NGC 4536, and NGC 4579), others have constant $r_{21}$ out to relatively large radii (i.e., NGC 3521, NGC 3621, NGC 4254, NGC 4321, and NGC 4826).

The top left-hand panel of Figure 4 shows the radial gradient of our galaxies combined. The general trend is consistent with previous studies (Y. Yajima et al. 2021; A. K. Leroy et al. 2022), which point out an elevated $r_{21}$ in the central 1-2 kpc of around $r_{21} = 0.6 - 0.7$. There is considerable variation between galaxies, however. Most notably, galaxies with SB morphology have a clearly decreasing trend, whereas SA galaxies maintain a relatively constant or shallow gradient in $r_{21}$. The outskirts of some of the barred galaxies (i.e., NGC 1512, NGC 3351, and NGC 4579) can reach as low as $r_{21} \sim 0.45$. LVG calculations of isothermal spherical clouds (e.g., S. Sakamoto et al. (1994)) indicate that such ratios are realized for gas around and below the critical density of $CO_{10}$ emission, under a range of temperatures and CO abundances.

The bimodal nature of the radial trend in $r_{21}$ is more pronounced in the top right-hand panel, where the radius has been normalized to the optical radius $R_{25} = D_{25}/2$ and $r_{21}$ is normalized to the value averaged over the central beam, i.e., out to a radii of $28''$. All SA galaxies and two in the SAB classification (i.e., NGC 3627 and NGC 4321) maintain a constant or very shallow gradient in $r_{21}$ out to $0.4R_{25}$, decreasing only by $\sim 5\%$. On the other hand, all SB galaxies and the remaining SABs (i.e., NGC 4536 and NGC 4579) quickly decrease in $r_{21}$ by 15% already before $0.3R_{25}$, reaching 20% before $0.4R_{25}$. The two categories can be clearly separated at 0.3 to $0.4R_{25}$.

In the bottom panel of Figure 4 we show only the barred galaxies (SAB and SB), where the radius is normalized to the characteristic length of the stellar bar, defined as where the amplitude of the $m = 2$ mode of the Fourier component in the near-IR image reaches a maximum (S. Dìaz-Garcìa et al. 2016). We see here again the bimodal trend (although less pronounced) where all SB galaxies and SAB galaxies except NGC 3627 and NGC 4321 decrease in $r_{21}$ by 10% at a radius corresponding the characteristic size of its stellar bar. Interestingly, NGC 3627 and NGC 4321 can be noted for their lopsided appearance, which could indicate that a simple radial trend would not be able to characterize the large scale variation in $r_{21}$.

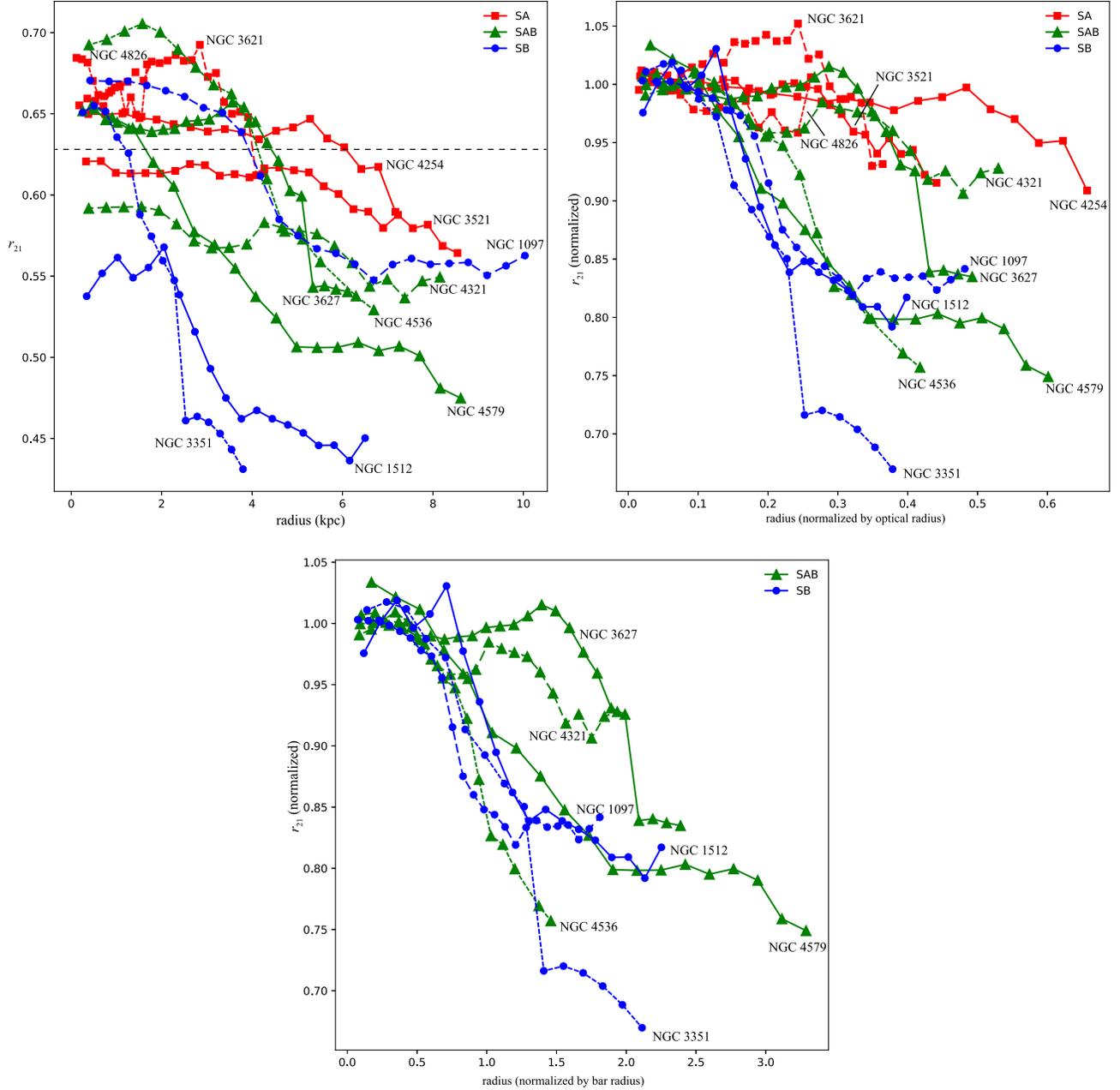

**Figure 4.** Top left-hand panel: Radial gradient of $r_{21}$ in the disk to the galactocentric radius. The dashed horizontal line corresponds to $r_{21} = 0.63$, the average value for all galaxies with the disk and outliers combined. Red squares, green triangles, and blue circles represent SA, SAB, and SB galaxies, respectively. Top right-hand panel: Same as top figure but with $r_{21}$ and radius normalized to the mean value in the central $28''$ and optical radius $R_{25}$, respectively. Bottom panel: Same as top right-hand panel, but for SAB and SB galaxies but with radius normalized to the characteristic radius of the bar, as given in Table 1.

We do not see low $r_{21}$ within the bar radius as found in individual sources (K. Muraoka et al. 2016; F. Egusa et al. 2022; F. Maeda et al. 2022, 2023; J. S. den Brok et al. 2023). This is likely because we lack the angular resolution to allow us to distinguish the rapid change in $r_{21}$ expected from the center, bar, and bar ends.

At radii well outside of the bars, even the galaxies with rapidly decreasing $r_{21}$ transition to a mode of more stable $r_{21}$ with a shallow gradient, much like the SA galaxies. In this regard, barred and nonbarred spiral galaxies both seem to maintain a relatively constant or shallow $r_{21}$ where the spiral structure prevails.

### 4.4. On the Use of a Single Value for $r_{21}$

Conversion of CO intensity to molecular mass involves the CO-to-$H_2$ conversion factor $X_{CO}$, which itself has a large uncertainty, depending on physical conditions such as metallicity or underlying molecular cloud populations (A. M. Lee et al. 2024). Considering this, one may argue that the variation in $r_{21}$ seen here does not significantly affect results obtained by observing only the $CO_{21}$ under the assumption of a constant $r_{21}$. However, the variations in $r_{21}$ we find are systematic with respect to galaxy morphology and radius, which can affect results in the literature. The change within galaxies is such that $r_{21}$ decreases by ~20% at large radii where the gas surface density is low in a group containing barred galaxies.

This can affect, for example, the power-law index in the Schmidt–Kennicutt (SK) relation. The SK relation is a positive correlation between the SFR surface density $\Sigma_{SFR}$ and the molecular gas surface density $\Sigma_{H2}$, characterized by a power-law index $N$ so that $\Sigma_{SFR} \propto \Sigma_{H2}^N$. A point of interest is whether the relation is linear ($N = 1$), indicating that molecular gas is converted to stars with fixed efficiency, or super-linear ($N > 1$), meaning that denser gas form stars more efficiently. Studies using $CO_{10}$ to trace molecular mass have commonly observed super-linear relations (S. Komugi et al. 2005; G. Liu et al. 2011; R. Momose et al. 2013), while those using higher-$J$ CO lines have derived tighter relations with $N \sim 1$ (S. Komugi et al. 2007; K. Muraoka et al. 2007; A. K. Leroy et al. 2008; F. Bigiel et al. 2008; D. Iono et al. 2009; A. Schruba et al. 2011; F. Bigiel et al. 2011; K. Muraoka et al. 2016) when assuming a constant ratio between the $J = 1 - 0$ and higher-$J$ emission. The value of $N$ can easily change with how the local SFR is estimated (G. Liu et al. 2011; R. Momose et al. 2013), or possibly on the evolutionary stage of the molecular clouds (S. Komugi et al. 2012) and if other physical environments are included in the relation in an extended form (Y. Shi et al. 2011; S. Rahmani et al. 2016; S. Komugi et al. 2018).

Y. Yajima et al. (2021) demonstrate a change in $N$ for a sample of galaxies when using $CO_{10}$, and find that the use of $CO_{21}$ and fixed $r_{21}$ would systematically underestimate $N$ because $r_{21}$ is higher in the denser parts of the molecular gas. In fact, of the seven galaxies in their sample (i.e., NGC 2798, NGC 2976, NGC 3351, NGC 4536, NGC 4569, NGC 4579, and NGC 5713) for which $N$ changed by more than ~20%, all but NGC 2976 are barred galaxies. Three galaxies of these seven overlap with our sample (i.e., NGC 3351, NGC 4536, and NGC 4579), and they all lie in the group with steeply decreasing $r_{21}$. Furthermore, galaxies in the group with shallow $r_{21}$ gradient which overlap with the Y. Yajima et al. (2021) sample (i.e., NGC 3521, NGC 3627, NGC 4254, and NGC 4321) only have a small change in $N$ of ~10% or less. This can be partially explained by our results which indicate that for barred galaxies $r_{21}$ decreases from the center where the CO is generally stronger to the outskirts where CO is weaker. A fixed $r_{21}$ would be characteristic of a value at a certain radius near the center. If we take $r_{21} = 0.7$ at the center, $r_{21} = 0.63$ as a global average, and $r_{21} = 0.5$ at large radii, then $\Sigma_{H2}$ at the center (with high $\Sigma_{H2}$) for a fixed $r_{21}$ would be overestimated with respect to the case of a varying $r_{21}$ by $\log \frac{0.7}{0.63} \sim 0.05$. Similarly, $\Sigma_{H2}$ at large radii (with low $\Sigma_{H2}$) for a fixed $r_{21}$ would be underestimated by $\log \frac{0.63}{0.5} \sim 0.1$. Since the typical range of $\Sigma_{H2}$ encountered per galaxy in previous observations is ~1 dex, and $\Sigma_{SFR}$ would not be changed, the decreased range in $\Sigma_{H2}$ using a radially varying $r_{21}$ would change $N = 1$ to $N = \frac{1}{1 - (0.05 + 0.1)} \sim 1.17$, which is an increase by ~20%. This is a very crude estimate but can nevertheless explain the large change in $N$ obtained by Y. Yajima et al. (2021) for barred galaxies.

The actual effect of an $r_{21}$ gradient on the SK index $N$ is likely intermixed with the effect coming from a radial gradient in the metallicity. Studies spanning a range of metallicities have shown the conversion factor $X_{CO}$ to increase at low metallicities (C. D. Wilson 1995; N. Arimoto et al. 1996; F. P. Israel 1997; A. K. Leroy et al. 2011; S. Komugi et al. 2011; A. D. Bolatto et al. 2013; L. K. Hunt et al. 2023; S. Komugi et al. 2023). Gas at large galactocentric radii typically have lower metallicities compared to the center, resulting in a radial gradient in $X_{CO}$ with slopes of ~0.5dex (A. D. Bolatto et al. 2013; K. M. Sandstrom et al. 2013; Q. Jiao et al. 2021). This corresponds to approximately a factor 3 change in molecular gas mass at a radius of $0.5R_{25}$, where our measurements reach. Thus, it seems that the effect of a metallicity gradient is larger than the effect of an $r_{21}$ gradient, but this does not take into account that the slope in the metallicity gradient may have morphology dependence. Spectroscopic studies in a number of spiral galaxies (M. B. Vila-Costas & M. G. Edmunds 1992; D. Zaritsky et al. 1994; I. Pessa et al. 2023) have shown that the metallicity gradient is shallow in barred galaxies (~20% decrease at $0.5R_{25}$) and steep in nonbarred galaxies (~60% decrease at $0.5R_{25}$). Depending on the choice of the metallicity-$X_{CO}$ relation, this could result in the effect of variation in $r_{21}$ (~20% $0.5R_{25}$ for the group containing barred galaxies) being comparable to or outweighing the effect of metallicity in terms of molecular gas mass estimation, strengthening the grounds for steeper $N$ in barred galaxies. For nonbarred galaxies, metallicity would be the dominant source of variation in measuring molecular mass.

## 5. Summary

This paper presented CO line ratio measurements between $^{12}CO(J = 2 - 1)$ and the $^{12}CO(J = 1 - 0)$ emission lines, both conducted with the ALMA TP array. Our targets are a complete subset of PHANGS-ALMA, Herschel KINGFISH, and Spitzer SINGS galaxies. The flux calibration accuracy offered by ALMA enables detailed investigation of the spatial and spectral variation of the CO line ratio $r_{21}$. Global average values of $r_{21} = 0.61 \pm 0.06$ are consistent with averages in the literature. By fitting tilted-ring models to the data cubes, we distinguish data that can be modeled by circular rotating disk from kinematic outliers. The outliers have a statistically larger mean value of $r_{21} = 0.69 \pm 0.10$ compared to the disk with $r_{21} = 0.62 \pm 0.08$. A comparison between different galaxy morphological types does not reveal systematic differences in the average values of $r_{21}$. We investigate the radial variation of $r_{21}$ in the disk component, and find a constant or shallow gradient for a subset of galaxies dominated by SA galaxies. For the remaining subset dominated by SB galaxies, we find a steeply declining slope characterized by the length of the stellar bar, at 20%–30% of the optical radius, beyond which the slope in $r_{21}$ becomes shallow. The correlation of $r_{21}$ with galaxy morphology and dynamics within galaxies shows that it is important to understand why and how the CO is excited at higher resolution and over a wider range of global galaxy characteristics.

## Acknowledgments

This paper makes use of the following ALMA data: ADS/JAO.ALMA#2022.1.00360.S, ADS/JAO.ALMA#2012.1.00650.S, ADS/JAO.ALMA#2015.1.00925.S, ADS/JAO.ALMA#2017.1.00886.L, and ADS/JAO.ALMA#2018.1.01651.S. ALMA is a partnership of ESO (representing its member states), NSF (USA), and NINS (Japan), together with NRC (Canada), NSTC and ASIAA (Taiwan), and KASI (Republic of Korea), in cooperation with the Republic of Chile. The Joint ALMA Observatory is operated by ESO, AUI/NRAO, and NAOJ. S.K. acknowledges support from the Yamada Research Foundation overseas research support. This work was conducted by S.K. as a visiting scholar at Stony Brook University and sabbatical researcher at National Astronomical Observatory of Japan. S.K was supported by the ALMA Japan Research Grant of NAOJ-ALMA-340. F.M. is supported by JSPS KAKENHI grant No. JP23K13142. F.E. is supported by JSPS KAKENHI grant No. 20H00172.

*Facilities:* ALMA, Herschel.

## ORCID iDs

Shinya Komugi ● https://orcid.org/0009-0007-2493-0973
Tsuyoshi Sawada ● https://orcid.org/0000-0002-0588-5595
Jin Koda ● https://orcid.org/0000-0002-8762-7863
Fumi Egusa ● https://orcid.org/0000-0002-1639-1515
Fumiya Maeda ● https://orcid.org/0000-0002-8868-1255
Akihiko Hirota ● https://orcid.org/0000-0002-0465-5421
Amanda M. Lee ● https://orcid.org/0000-0001-8254-6768

## References

Anand, G. S., Lee, J. C., Van Dyk, S. D., et al. 2021, MNRAS, 501, 3621
Arimoto, N., Sofue, Y., & Tsujimoto, T. 1996, PASJ, 48, 275
Athanassoula, E., & Bureau, M. 1999, ApJ, 522, 699
Bigiel, F., Leroy, A., Walter, F., et al. 2008, AJ, 136, 2846
Bigiel, F., Leroy, A. K., Walter, F., et al. 2011, ApJL, 730, L13
Bolatto, A. D., Wolfire, M., & Leroy, A. K. 2013, ARA&A, 51, 207
Braine, J., & Combes, F. 1992, A&A, 264, 433
Braine, J., Combes, F., Casoli, F., et al. 1993, A&AS, 97, 887
Buta, R. J., Sheth, K., Athanassoula, E., et al. 2015, ApJS, 217, 32
CASA Team, Bean, B., Bhatnagar, S., et al. 2022, PASP, 134, 114501
Crosthwaite, L. P., & Turner, J. L. 2007, AJ, 134, 1827
den Brok, J. S., Chatzigiannakis, D., Bigiel, F., et al. 2021, MNRAS, 504, 3221
den Brok, J. S., Leroy, A. K., Usero, A., et al. 2023, MNRAS, 526, 6347
de Vaucouleurs, G., de Vaucouleurs, A., Corwin, H. G., J., et al. 1991, Third Reference Catalogue of Bright Galaxies (New York: Springer)
Dìaz-Garcìa, S., Salo, H., Laurikainen, E., & Herrera-Endoqui, M. 2016, A&A, 587, A160
Di Teodoro, E. M., & Fraternali, F. 2015, MNRAS, 451, 3021
Egusa, F., Gao, Y., Morokuma-Matsui, K., Liu, G., & Maeda, F. 2022, ApJ, 935, 64
Falgarone, E., Panis, J. F., Heithausen, A., et al. 1998, A&A, 331, 669
Farren, G. S., Partridge, B., Kneissl, R., et al. 2021, ApJS, 256, 19
Freedman, W. L., Madore, B. F., Gibson, B. K., et al. 2001, ApJ, 553, 47
Gierens, K. M., Stutzki, J., & Winnewisser, G. 1992, A&A, 259, 271
Goldreich, P., & Kwan, J. 1974, ApJ, 189, 441
Handa, T., Hasegawa, T., Hayashi, M., et al. 1993, in AIP Conf. Ser. 278, Back to the Galaxy, ed. S. S. Holt & F. Verter (Melville, NY: AIP), 315
Hatchfield, H. P., Sormani, M. C., Tress, R. G., et al. 2021, ApJ, 922, 79
Hunt, L. K., Belfiore, F., Lelli, F., et al. 2023, A&A, 675, A64
Iguchi, S., Morita, K.-I., Sugimoto, M., et al. 2009, PASJ, 61, 1
Iono, D., Wilson, C. D., Yun, M. S., et al. 2009, ApJ, 695, 1537
Israel, F. P. 1997, A&A, 328, 471
Jiao, Q., Gao, Y., & Zhao, Y. 2021, MNRAS, 504, 2360
Kamazaki, T., Okumura, S. K., Chikada, Y., et al. 2012, PASJ, 64, 29
Keenan, R. P., Marrone, D. P., & Keating, G. K. 2024a, arXiv:2409.03963
Keenan, R. P., Marrone, D. P., Keating, G. K., et al. 2024b, ApJ, 975, 150
Kennicutt, R. C., J., Armus, L., Bendo, G., et al. 2003, PASP, 115, 928
Kennicutt, R. C., Calzetti, D., Aniano, G., et al. 2011, PASP, 123, 1347
Koda, J., Hirota, A., Egusa, F., et al. 2023, ApJ, 949, 108
Koda, J., Sawada, T., Sakamoto, K., et al. 2020, ApJL, 890, L10
Koda, J., Scoville, N., Hasegawa, T., et al. 2012, ApJ, 761, 41
Koda, J., & Wada, K. 2002, A&A, 396, 867
Komugi, S., Inaba, M., & Shindou, T. 2023, PASJ, 75, 1337
Komugi, S., Kohno, K., Tosaki, T., et al. 2007, PASJ, 59, 55
Komugi, S., Miura, R. E., Kuno, N., & Tosaki, T. 2018, PASJ, 70, 48
Komugi, S., Sofue, Y., Nakanishi, H., Onodera, S., & Egusa, F. 2005, PASJ, 57, 733
Komugi, S., Tateuchi, K., Motohara, K., et al. 2012, ApJ, 757, 138
Komugi, S., Yasui, C., Kobayashi, N., et al. 2011, PASJ, 63, L1
Kuno, N., Sato, N., Nakanishi, H., et al. 2007, PASJ, 59, 117
Lang, P., Meidt, S. E., Rosolowsky, E., et al. 2020, ApJ, 897, 122
Lee, A. M., Koda, J., Hirota, A., Egusa, F., & Heyer, M. 2024, ApJ, 968, 97
Leroy, A. K., Bolatto, A., Gordon, K., et al. 2011, ApJ, 737, 12
Leroy, A. K., Rosolowsky, E., Usero, A., et al. 2022, ApJ, 927, 149
Leroy, A. K., Schinnerer, E., Hughes, A., et al. 2021, ApJS, 257, 43
Leroy, A. K., Walter, F., Brinks, E., et al. 2008, AJ, 136, 2782
Liu, G., Koda, J., Calzetti, D., Fukuhara, M., & Momose, R. 2011, ApJ, 735, 63
Maeda, F., Egusa, F., Ohta, K., Fujimoto, Y., & Habe, A. 2023, ApJ, 943, 7
Maeda, F., Egusa, F., Ohta, K., et al. 2022, ApJ, 926, 96
Maeda, F., Ohta, K., Fujimoto, Y., Habe, A., & Ushio, K. 2020, MNRAS, 495, 3840
Makarov, D., Prugniel, P., Terekhova, N., Courtois, H., & Vauglin, I. 2014, A&A, 570, A13
McQuinn, K. B. W., Skillman, E. D., Dolphin, A. E., Berg, D., & Kennicutt, R. 2017, AJ, 154, 51
Momose, R., Koda, J., Kennicutt, R. C., J., et al. 2013, ApJL, 772, L13
Muraoka, K., Kohno, K., Tosaki, T., et al. 2007, PASJ, 59, 43
Muraoka, K., Sorai, K., Kuno, N., et al. 2016, PASJ, 68, 89
Nishimura, A., Tokuda, K., Kimura, K., et al. 2015, ApJS, 216, 18
Oka, T., Hasegawa, T., Hayashi, M., Handa, T., & Sakamoto, S. 1998, ApJ, 493, 730
Pessa, I., Schinnerer, E., Sanchez-Blazquez, P., et al. 2023, A&A, 673, A147
Rahmani, S., Lianou, S., & Barmby, P. 2016, MNRAS, 456, 4128
Regan, M. W., Sheth, K., & Vogel, S. N. 1999, ApJ, 526, 97
Rogstad, D. H., Lockhart, I. A., & Wright, M. C. H. 1974, ApJ, 193, 309
Sakamoto, S., Hasegawa, T., Handa, T., Hayashi, M., & Oka, T. 1997, ApJ, 486, 276
Sakamoto, S., Hayashi, M., Hasegawa, T., Handa, T., & Oka, T. 1994, ApJ, 425, 641
Sandstrom, K. M., Leroy, A. K., Walter, F., et al. 2013, ApJ, 777, 5
Sawada, T., Chang, C.-S., Francke, H., et al. 2021, PASP, 133, 034504
Sawada, T., Hasegawa, T., Handa, T., et al. 2001, ApJS, 136, 189
Schruba, A., Leroy, A. K., Walter, F., et al. 2011, AJ, 142, 37
Schwab, F. R. 1980, Proc. SPIE, 231, 18
Schwab, F. R. 1984, in Indirect Imaging. Measurement and Processing for Indirect Imaging, ed. J. A. Roberts (Cambridge: Cambridge Univ. Press), 333
Scoville, N. Z., & Solomon, P. M. 1974, ApJL, 187, L67
Shi, Y., Helou, G., Yan, L., et al. 2011, ApJ, 733, 87
Sorai, K., Kuno, N., Muraoka, K., et al. 2019, PASJ, 71, S14
Sormani, M. C., Barnes, A. T., Sun, J., et al. 2023, MNRAS, 523, 2918
Tress, R. G., Sormani, M. C., Glover, S. C. O., et al. 2020, MNRAS, 499, 4455
Tully, R. B., Courtois, H. M., Dolphin, A. E., et al. 2013, AJ, 146, 86
Tully, R. B., Courtois, H. M., & Sorce, J. G. 2016, AJ, 152, 50
van der Tak, F. F. S., Black, J. H., Schöier, F. L., Jansen, D. J., & van Dishoeck, E. F. 2007, A&A, 468, 627
Vila-Costas, M. B., & Edmunds, M. G. 1992, MNRAS, 259, 121
Wilson, C. D. 1995, ApJL, 448, L97
Yajima, Y., Sorai, K., Miyamoto, Y., et al. 2021, PASJ, 73, 257
Zaritsky, D., Kennicutt, R. C., Jr., Huchra, J. P., et al. 1994, ApJ, 420, 87